\documentclass{article}

\usepackage{arxiv}

\usepackage[utf8]{inputenc} 
\usepackage[T1]{fontenc}    
\usepackage{hyperref}       
\usepackage{url}            
\usepackage{booktabs}       
\usepackage{amsfonts}       
\usepackage{nicefrac}       
\usepackage{microtype}      
\usepackage{lipsum}		
\usepackage{graphicx}

\usepackage{natbib}

\usepackage{amsmath}
\usepackage{array}

\title{A Review of Measurement Error Bias-Correction in Nutritional Epidemiology}

\author{ \href{https://orcid.org/0000-0001-7431-1619}{\includegraphics[scale=0.06]{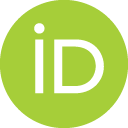}\hspace{1mm}Huimin Peng}\thanks{This manuscript is composed from my written preliminary exam manuscript of my PhD program. I received PhD in statistics from North Carolina State University. Great thanks to anonymous professors who leave helpful comments on this manuscript and recommends useful writing skills to help me improve this manuscript.} \\
		\texttt{hpeng2@ncsu.edu}\\
	\texttt{penghuimin@inspur.com}\\
	\texttt{peng.huimin.pennie@gmail.com} \\
}

\begin{document}
\maketitle

\begin{abstract}
This article reviews bias-correction models for measurement error of exposure variables in the field of nutritional epidemiology. 
Measurement error usually attenuates estimated slope towards zero. Due to the influence of measurement error, inference of parameter estimate is conservative and confidence interval of the slope parameter is too narrow. Bias-correction in estimators and confidence intervals are of primary interest. We review the following bias-correction models: regression calibration methods, likelihood-based models, missing data models, simulation-based methods, nonparametric models and sampling based procedures. 
\end{abstract}


\section{Introduction}
\label{Introduction}

Nutritional epidemiology studies focus on unveiling the relationship between exposure factors and disease status. Researchers believe that there is a connection between dietary habits and the probability of being infected with disease or deteriorating into cancer. Typical examples of exposure variables are nutrient intake, dietary intake and total energy intake \citep{b25,b22}.

Exposure factors can be investigated through a long-term survey FFQ (food frequency questionnaire) \citep{b3}. FFQ surveys the subjects on consumption frequencies and  portion sizes of different primary food over a period of time usually six months or a whole year \citep{b20}. Researchers then estimate  nutrient intake or energy intake from  questionnaires \citep{b27}. It is relatively easy and of low cost to obtain FFQ. Sample size of FFQ is usually very large. However, FFQ is subject to a substantial possibility of being contaminated with measurement error, since there is randomness when people recall food consumption history, and  different food pool included in FFQ may not be representative of everyone's dietary habit \citep{b3, b51,b22}. 

In the presence of measurement error,  estimated effect of exposure factors on  disease status is biased \citep{b27}.  Direction of bias is unknown and may be upward or downward. Bias depends on the assumptions on measurement error and the correlation structure between exposure factors. Estimated confidence interval is also biased \citep{b2,b15}. Considering bias in estimates and confidence intervals,  measurement error correction models are adopted to derive unbiased point estimates, and accurate inference on unknown parameters. Researchers understand the association between nutrient intake and disease status better based upon unbiased estimation techniques.  

Other methods of investigating exposure factors include 24-hour recall (24HR), food record(FR), 7-day diary (7DD) and biomarker \citep{b22}. These methods provide more accurate dietary information than FFQ but are also more expensive \citep{b23}. Due to the cost-efficiency of study design, these methods are applied on a smaller sample or a sub-sample of FFQ \citep{b21}. In real-life applications, precise records are used as validation or calibration studies to evaluate measurement error in FFQ \citep[chap.~4]{b41}. They are regarded as the "golden standards" or the reference measurements to study the relationship between unobserved true exposure and observed predictor measured with error \citep{b20,b22,b38}. "Golden standard" assumes that measurement errors are inconsequential in precise records. The premise for applying validation study is that measurement error in the validation study is independent of that in FFQ \citep{b22}. In addition, it is assumed that relationship between true and observed exposure on the precise validation data is the same as on the contaminated primary data \citep{b34}.

Data collected in the nutritional epidemiology studies can be grouped into five classes, as in \citet{b33}. Primary data includes disease status, observed predictors measured with error and confounding predictors measured without error. Internal validation data is comprised of disease status, confounding factors, true exposure and observed covariates with measurement error. External validation data contains only the last two variables in the internal validation data. Validation data is used to estimate the relationship between observed and true exposure. Internal reliability data consists of disease status and repeated measurements on the observed predictor. External reliability data includes only the repeated measurements of observed predictor. We can estimate the variance of measurement error from the reliability data. Primary data and at least one of the validation and the reliability data must be present to correct the bias induced by measurement error \citep{b33}. 

\subsection{Definition}
\label{Definition}

Nondifferential measurement error is independent of outcome variables such as disease status. Conversely differential measurement error is dependent of outcome variable that is, the distribution of measurement error varies at different disease statuses \citep{b17}. Most statistical models assume nondifferential measurement error and that disease status $D$ and observed predictor $Z$ measured with error are conditionally independent given true exposure $X$ and confounding factors $W$, that is $D\perp Z|(X, W)$ (conditional independence assumption). 

Random measurement error is due to imprecision and fluctuates around zero \citep{b20, b17}. It can be measured by taking repeated measurements on the same individual \citep[chap.~12]{b41}. Systematic measurement error leads to bias on average \citep{b20}. For instance, subjects may consistently over-recall food consumption frequency or consistently under-recall it. For different individuals, systematic measurement errors may be in different directions and of different sizes. It is generally more difficult to evaluate systematic measurement error than  random measurement error, as in \citet[chap.~12]{b41}. 

Denote $\epsilon$ to be measurement error, then additive measurement error follows that $Z=\alpha^\prime+X+\epsilon$, where $\alpha^\prime$ is the systematic measurement error and $\epsilon$ is the random measurement error with mean $0$. Multiplicative measurement error follows that $Z=X\epsilon$, where $\epsilon$ is the random measurement error with mean $1$ \citep{b34,b40}. In most measurement error models, measurement error is assumed to be additive. By taking the logarithm transformation of all variables, multiplicative measurement error should be additive as well \citep{b3}.

\citet{b2} summarizes measurement error in nutritional epidemiology to be in four types: within-subject random error, within-subject systematic error, between-subject random error, and between-subject systematic error. Within-subject random error fluctuates around zero for each individual. Within-subject systematic error brings individual bias to the variable on average in different directions and of different sizes. Between-subject random error fluctuates around zero on average for each subject enrolled in the study. Between-subject systematic error brings bias to the variable on average for all subjects. \citet{b20} mentions that measurement error can be classified into person-specific or food-specific error. Measurement error can be between-person or within-person. It can also be within-food or between-food since every food consumption is measured with error in FFQ. 

\subsection{Complications}
\label{Complications}

In the univariate measurement error model where measurement error is assumed to be nondifferential and where conditional independence assumption holds, measurement error attenuates the estimated parameter towards its null value \citep{b6,b3}. As the variability ratio between measurement error and true exposure increases, bias size also increases. Naive confidence interval estimate without bias correction is narrow and conservative. When measurement error is differential or depends upon true exposure, bias may be upward or downward depending upon the correlation between exposure factors. Bias size is also related to whether the study design is randomized and balanced \citep{b34}.

Both correction methods in linear and  nonlinear models are applicable in nutritional epidemiology. For categorical outcome variable, bias correction in nonlinear regression such as logistic regression is of particular interest \citep{b1}. For energy-adjusted model, bias correction in multiple linear regression with complex correlation structure is of concern \citep{b6}.  

When testing null hypothesis that the true exposure has no effect on disease status,
measurement error leads to loss in testing power \citep{b34,b17}. Uncorrected power function will generally be greater and sample size required in the experimental design is usually under-estimated \citep{b30}. For tests based upon contaminated data with measurement error, sample size required to achieve the same power is greater \citep{b6,b2,b3}. Under the framework of regression calibration model with Gaussian exposure, sample size for achieving the same statistical power is proportional to the quantity: $1/(\lambda^2\sigma^2_Z)$, where $\lambda$ is the attenuation factor and $\sigma^2_Z$ is the variance of observed exposure \citep{b22}. If measurement error variance is large, the attenuation factor tends to be small and the required sample size is very large to achieve the same power.  

Researchers in nutritional epidemiology are constantly searching for more accurate measurements of nutrient intake \citep{b41}. Statisticians are developing more measurement error models to relax assumptions and to improve the performance of bias-corrected estimators. Extensions of bias-correction models are from linear \citep{b4} to nonlinear \citep{b15} or partially linear models \citep{b29}. Bias-correction models include regression calibration method \citep{b1,b11}, quasi-likelihood approximation method \citep{b33,b35}, Bayesian nonparametric and semi-parametric model \citep{b8,b37,b7,b29} and simulation extrapolation method \citep{b16,b18}. 

Structure of our survey article is as follows.
Section \ref{Methodology} reviews main bias-correction methods on measurement error in nutritional epidemiology.
Section \ref{regression-calibration} presents regression calibration methods.
Section \ref{likelihood-based-methods} surveys likelihood-based models.
Section \ref{missing-data-perspective} reviews bias-correction methods which regard measurement error as missing data.
Section \ref{simulation-based-methods} summarizes simulation-based bias-correction procedure.
Section \ref{nonpar-methods} presents nonparametric and semi-parametric methods.
Section \ref{simulation} conducts a simulation study to compare performances of bias-correction methods.
Section \ref{discussion} discusses current methods and future research directions.

\section{Methodology}
\label{Methodology}

\subsection{Formulation and Assumptions}
\label{Assumptions}

Denote $X$ to be true exposure and $Z$ to be observed exposure or surrogate, measurement error model is
\begin{eqnarray}
Z_i=\alpha^\prime+c(X_i,\eta)+\epsilon_i,~i=1,\cdots ,N, \nonumber
\end{eqnarray}
where $\alpha^\prime$ is an unknown parameter, $\epsilon_i$ is the measurement error, $N$ is the number of subjects enrolled in the study and $c(\cdot,\cdot)$ is a known function of $X_i$ \citep{b33}. 
Classical assumptions on measurement error model are as follows \citep{b6}.
\begin{enumerate}
	\item Measurement error $\epsilon_i$ is independent of true exposure.
	\item Measurement errors of different exposure factors are uncorrelated.
	\item Measurement error is homoscedastic.
\end{enumerate}
Assumptions imposed on the conditional distributions of $\epsilon|X$ can be further relaxed. 

On the other hand, Berkson measurement error model \citep{b34} is formulated as 
\begin{eqnarray}
X_i=\alpha^\prime+c^*(Z_i,\eta)+\epsilon_i,~i=1,\cdots ,N, \nonumber
\end{eqnarray}
where $c^*(\cdot,\cdot)$ is a known function of observed exposure $Z_i$. Berkson model focuses on the conditional distributions of $\epsilon|Z$ and $X|Z$ rather than $\epsilon|X$ \citep{b17}.

In functional model,
true exposure $X$ is regarded as fixed and nonrandom. Conversely, structural model assumes random true exposure \citep{b34,b17}. We can construct a sufficient statistic of $X$ and condition the likelihood function on the sufficient statistic, then a structural model is transformed into a functional model. After a sufficient statistic is derived, conditional maximum likelihood estimate is used as a bias-corrected estimator \citep{b46}. 


Confounding factors correlated with both disease status $D$ and observed exposure $Z$ are of concern in nutritional epidemiology study \citep{b20}. Furthermore, measurement errors may be correlated with true exposure and outcome variable. Measurement error in exposure factors may be correlated. Joint distribution of observed and true exposure is usually non-Gaussian and highly skewed. Random within-subject measurement error may be heteroscedastic \citep{b13}. Most statistical models only consider random within-subject error since correction of random measurement error only requires repeated measurements on the same subject, while correcting for systematic measurement error relies on  replicate validation studies \citep[chap.~12]{b41}.

\subsection{Regression Calibration Methods}
\label{regression-calibration}

Regression calibration model uses a contaminated primary dataset and a precise validation dataset. Contaminated exposure factors are precisely measured on the validation data. By calibrating contaminated exposure upon precisely measured exposure on the validation data, we can estimate a mapping between contaminated exposure and precisely measured exposure. We assume that this mapping can be extrapolated from the small validation sample to all subjects in large primary data. 
Then we can use calibrated exposure factors in primary data to analyze the relation between disease and nutrition intake.

\citet{b1} assumes that measurement error is nondifferential and independent of true exposure. Only random and systematic within-subject measurement errors are considered. It is common to use logistic regression to study the relation between  disease status and exposure. \citet{b1} proposes two bias-correction methods for slope parameter in logistic regression when exposure factors are measured with error. The model consists of two submodels: logistic regression for main data and linear regression for validation data. We denote $D$ to be the disease status, $Z$ as the observed exposure measured with error, and $X$ to be the true exposure. For simplicity, $D$ is binary. Regression calibration model is
\begin{eqnarray}
\mbox{logit}\{P(D|X)\}&=&\alpha+\beta X,\nonumber\\
X&=&\alpha^\prime +\lambda Z+\epsilon,\nonumber
\end{eqnarray}
where $\epsilon\sim N(0,\sigma^2)$ is the Gaussian measurement error, $\alpha$, $\beta$, $\alpha^\prime$, $\lambda$ and $\sigma$ are the unknown parameters. Logit function is $\mbox{logit}(x)=\log(x/(1-x))$. Parameter $\lambda$ is the attenuation factor and $\lambda$ is usually less than $1$. In some cases, $\lambda$ can be greater than $1$ as well \citep{b22}.

\citet{b1} requires that the dataset should include a contaminated primary study and a precise small-scale validation study. First, we review a regression calibration model based upon linear approximation \citep{b11}. Steps in the procedure are as follows.
\begin{enumerate}
	\item We use ordinary least squares to regress true exposure $X$ on observed exposure $Z$ in the validation data and obtain estimates $\hat{\alpha}^\prime$ and $\hat{\lambda}$.
	\item Then we use the predicted exposure factors $\tilde{X}=\hat{\alpha}^\prime+\hat{\lambda} Z$ in the main data and fit a logistic regression of disease status $D$ on $\tilde{X}$ to obtain  a bias-corrected slope estimate $\hat{\beta}_1$. 
\end{enumerate}
Under the assumption that measurement error in the validation data is independent of true exposure and measurement error in primary study, bias-corrected slope estimator from regression calibration method is consistent \citep{b11}.

After calibration of measurement-error-contaminated exposure factors on precise validation data, predicted exposure factors are free of measurement error.
This bias-correction process extrapolates measurement error pattern on the small-scale validation data to the large primary data. 
For the extrapolation to hold,
it is vital to assume that measurement error distribution on the small precisely calibrated sample is the same as on the large primary data. 

For other more complex structural models, the idea behind regression calibration method is the same, that is, to replace the contaminated observed exposure in primary study with predicted exposure factors from a regression in precise validation data.
We can fit a logistic regression of disease status $D$ on the predicted exposure $\tilde{X}$ to obtain the regression calibration estimate of slope parameter \citep{b38}. For more than one exposure factors measured with error, we can find the corresponding predicted exposures from calibration on the validation study and conduct the regression of disease status $D$ on predicted exposures. 

A different perspective of regression calibration estimator is described in \citet{b1}. This procedure obtains the same point estimator thus it is also a regression calibration method. Steps in the procedure are as follows.
\begin{enumerate}
	\item We can fit a logistic regression of disease status $D$ on observed exposure $Z$ in the main data and calculate slope estimate $\hat{\beta}$.
	\item We may use ordinary least squares estimation and linear regression of reference measurements $X$ on FFQ measurements $Z$ in the validation data to obtain an attenuation factor estimate $\hat{\lambda}$. 
	\item Regression calibration estimator for $\beta$ is $\hat{\beta}_2=\hat{\beta}/\hat{\lambda}$.
	\item  The standard error of estimator is derived through Delta method assuming that $\hat{\beta}$ and $\hat{\lambda}$ are independent. 
\end{enumerate}

Estimation of the attenuation factor under different correlation structures between exposure factors in regression calibration measurement error model is of particular interest in nutritional epidemiology \citep{b20}. After estimating the attenuation factor, regression calibration corrected estimator is the biased naive estimator over estimated attenuation factor. Attenuation factor is usually between $0$ and $1$. Naive estimator is usually biased toward null value $0$. Bias-corrected estimator is closer to the true value than naive estimator. \citet{b51} investigates the estimation of attenuation factor when systematic bias in the primary data and the validation data are correlated. The standard error of regression calibration estimator derived with Delta method is severely under-estimated. Bootstrap scheme proposed in \citet{b32} is a better way to estimate variance in the regression calibration estimator. 

Estimators proposed in \citet{b1} are derived under stricter assumptions. It requires that random measurement error should be Gaussian and independent of true exposure and disease status. It also requires conditional independence assumption to be true. It does not consider the effect of confounding variable $W$ in regression model. 
\citet{b6} studies energy-adjusted model in nutritional epidemiology and uses measurement error correction methods in multiple linear regression. It posits model to be 
\begin{eqnarray}
E(Y|X)=\alpha+\beta_1 X_1+\beta_2 X_2,\nonumber
\end{eqnarray}
where $Y$ is the response variable and $X_1$ and $X_2$ are the true exposure factors. The observed predictors $Z_1$ and $Z_2$ are measured with error and $X_1$ and $X_2$ are true values of $Z_1$ and $Z_2$ respectively. 
Regression calibration model indicates that we apply a regression model to calibrate the relationship between true exposure $X$ and  observed exposure $Z$. \citet{b6} formulates the bias-correction model as
$X=\alpha^\prime+\lambda_1 Z_1+\lambda_2 Z_2+\epsilon$,
where $\alpha^\prime$, $\lambda_1$ and $\lambda_2$ are the unknown parameters and $\epsilon$ is the measurement error. It assumes that measurement error $\epsilon$ has a constant variance, that $\epsilon$ is independent of $Z_1$ and $Z_2$ and that the joint distribution of $Z_1$, $Z_2$ and $X$ is Gaussian. \citet{b6} provides an insight into bias induced by measurement error. We fit the model
\begin{eqnarray}
Y=\gamma_0+\gamma_1 Z_1+\gamma_2 Z_2+\varepsilon ,\nonumber
\end{eqnarray}
where $\gamma_0$, $\gamma_1$ and $\gamma_2$ are the unknown parameters and $\varepsilon$ is the random error, then asymptotic expectation of ordinary least squares estimate $\hat{\gamma_1}$ is
\begin{eqnarray}
E(\hat{\gamma_1})=A\beta_1 +C\beta_2,\nonumber
\end{eqnarray}
where $A$ and $C$ are two conformable constant matrices. Matrix $A$ reflects the bias of $\hat{\gamma_1}$ due to measurement error of $Z_1$ and $A$ is an attenuation factor. Matrix $C$ represents the bias of $\hat{\gamma_1}$ caused by measurement error of $Z_2$. Matrices $A$ and $C$ are only relevant to the correlation structure between $Z_1$, $Z_2$, $X_1$ and $X_2$. That is, correlation structure between predictors in the model and measurement errors has a crucial effect on the bias caused by measurement error. Bias direction is indefinite under complex correlation structure. In a model with multiple exposures, bias on the estimated effect is subject to both attenuation effect ($A$) of its own measurement error and contamination effect ($C$) from measurement errors of other exposures in the model. 

\citet{b32} describes how to use bootstrap procedures in regression calibration method.
The steps are as follows.
\begin{enumerate}
	\item Get a bootstrap sample with replacement from the validation data.
	\item Estimate attenuation factor using bootstrap sample.
	\item Obtain a bootstrap sample with replacement from the primary data.
	\item Estimate bias-corrected estimator $\hat{\beta}$ with disease status and predicted exposure factors in the bootstrap sample.
	\item Repeat steps above for a sufficiently large number of iterations and obtain a vector of bias-corrected estimates. 
\end{enumerate}     
Bootstrap procedure can be used to provide a better estimate of the standard error for  regression calibration corrected estimator \citep{b38}.

\subsection{Likelihood-Based Methods}
\label{likelihood-based-methods}

Conditional independence assumption and Gaussian measurement error assumption may not hold. Quasi-likelihood approach does not require measurement error to be Gaussian. No distributional assumption is imposed upon measurement error. 

Another framework for deriving bias-corrected parameter estimate for measurement error contaminated data is to apply maximum likelihood approach. 
\citet{b1} suggests using an approximation of likelihood function $Pr(D|Z)$ to derive a bias-corrected estimator. 
Steps are as follows. 
\begin{enumerate}
	\item Fit a logistic regression of disease status $D$ on observed exposure $Z$ in the main data and obtain estimates $\hat{\alpha}$ and $\hat{\beta}$. 
	\item Then compute likelihood-based bias-corrected estimator of $\beta$.
	\item The standard error of corrected estimator is also derived with Delta method and explicitly expressed.
\end{enumerate}
Measurement error variance is estimated from the validation study.
\citet{b15} presents three approximate estimators in logistic regression with measurement error in exposure factors. In nutritional epidemiology, a common situation is that sample size of the validation study is much smaller than primary study. Small sample size is an obstacle in obtaining an efficient estimate of measurement error variance. 

Quasi-likelihood approach only specifies models for the first and second moment of disease status distribution conditional on exposure factors ($D|X,W$). Approximation may be applied to the likelihood function first and then a bias-corrected estimate is computed from the approximate score function. On the other hand, approximation may be applied to the corrected estimate directly. Quasi-likelihood method does not specify measurement error distribution \citep{b33}. \citet{b35} computes quasi-likelihood score function and derives corrected estimators based on conditions imposed upon $\epsilon|Z$ distribution under the framework of Berkson model. Quasi-likelihood approach is also used to correct the power function in hypothesis testing contaminated with measurement error.  \citet{b30} develops bias-correction methods for the power function using quasi-likelihood and generalized score test under fixed alternatives. Quasi-likelihood estimators reduce computation complexity \citep{b34}.

\subsection{Missing Data Perspective}
\label{missing-data-perspective}

Conditional independence assumption and Gaussian error assumption may not hold. Quasi-likelihood approach does not require measurement error to be Gaussian. 
Regarding measurement error as missing data removes the conditional independence assumption.
Bias-correction model proposed in \citet{b9} regards true exposure $X$ as missing data in the primary data and develops a measurement error estimation technique based on expectation-maximization algorithm (EM). It focuses only upon the main data and assumes that the magnitude of measurement error is already known. It considers including confounding covariates $W$ which are not affected by measurement error in the model. Confounding factors $W$ may be correlated with both disease status $D$ and true exposure $X$. For confounding factors $W$ in the model, \citet{b1} recommends 
\begin{enumerate}
	\item first regressing observed exposure $Z$ and disease status $D$ respectively on confounding $W$,
	\item then regressing residuals of $D$ on residuals of $Z$ to obtain parameter estimate.
\end{enumerate}
\citet{b9} assumes that parametric distribution of disease status given true exposure $D|X$ is in the exponential family
\begin{eqnarray}
f(d_i|x_i;\beta)=\exp\{d_i(w_i^\prime\beta_1+x_i^\prime\beta_2)-b(w_i^\prime\beta_1+x_i^\prime\beta_2)+h(d_i)\}, i=1,\cdots ,N, \nonumber
\end{eqnarray}
where $d_i$, $x_i$ and $w_i$ are the realizations of $D_i$, $X_i$ and $W_i$ respectively, and $v^\prime$ stands for the matrix transpose of $v$. It assumes that conditional independence assumption holds and that measurement error is nondifferential. Random variables $D_i$, $Z_i$, $X_i$ of different subjects are independent. It also assumes that $Z_i|X_i=x_i\sim N(x_i,\Omega_m)$, that $X_i\sim N(x,\Omega_x)$, and that covariates $W,X$ and measurement error are jointly Gaussian. It assumes that measurement error variance $\Omega_m$ is known or has an efficient estimator $S_m$, so that it does not consider modeling the validation data. Unknown parameter in the model is denoted by  $\xi=(\beta_1,\beta_2,x,\Omega_x)$.

In Expectation-Maximization (EM) algorithm, complete data is $(D,Z,X,S_m)$ and observed data is $(D,Z,S_m)$ \footnote{Only large contaminated primary data is considered here.}. Expectation step and maximization step are respectively
\begin{eqnarray}
Q(\xi|\xi^{(t)})&=&E\{\log f(D,Z,X,S_m;\xi)|D,Z,S_m;\xi^{(t)}\},\nonumber\\
\xi^{(t+1)}&=&\mbox{argmax}_{\xi} ~ Q(\xi|\xi^{(t)}).\nonumber
\end{eqnarray}
Decomposition of log-likelihood is
\begin{eqnarray}
\log f(D,Z,X,S_m;\xi)&=&\sum\log f(D_i|X_i;\beta)+\sum\log f(X_i;x,\Omega_x)+\sum\log f(Z_i|X_i;\Omega_m)\nonumber\\
&+&\sum\log f(S_m;\Omega_m).\nonumber
\end{eqnarray}
Regression calibration model only considers distributions of $D|X$ and $Z|X$ which are the first and third items in the complete log-likelihood function. It iterates EM algorithm to estimate unknown parameters. In nutritional epidemiology study where we can estimate measurement error variance efficiently from the validation or reliability study, EM algorithm is a feasible bias-correction method.  

Multiple imputation method \citep{b39} requires that a precise validation study should be conducted and that the sample in validation study should be a sub-sample of the large primary study. Similar to EM algorithm, multiple imputation method views measurement error as missing data. It assumes that the true exposure is missing at random in the primary data. For observations which lie both in the primary and the validation study, true exposure is regarded as observed exposure. But for observations which only appear in the primary study and not in the validation study, true exposure is viewed as missing data. 

From validation study, the mapping between true exposure and observed exposure is estimated. Multiple imputation of true exposure is based upon the extrapolation of this estimated mapping. Each imputation produces a primary study with all true exposure factors filled with imputed observations. We can fit a logistic regression of disease status on the filled true exposure ($D\sim X^*$) factors. Rubin's estimate from multiple imputation procedure is equal to the average of all estimates. Estimate variance is the summation of between-imputation variance and within-imputation variance. Multiple imputation estimate is referred to as MIME \citep{b39}. MIME is approximately unbiased. The coverage of confidence interval is around the right coverage. The performance of MIME depends on the sample size of primary study and the proportion of subjects that are validated. 

\subsection{Simulation-Based Methods}
\label{simulation-based-methods}

SIMEX \citep{b16} is a simulation-based method to correct for measurement error. The assumption is that measurement error variability is known or can be efficiently estimated from the validation or reliability study. The procedure includes the following steps.
\begin{enumerate}
	\item Add an extra measurement error to the predictor measured with error.
	\item Estimate the parameter of interest using disturbed main data.
	\item Repeat steps above at different values of measurement error variance and computing the parameter estimate.
	\item Calibrate the relationship between  variance of extra measurement error and parameter estimate.
	\item Extrapolate the relationship to cases with no measurement error and obtain a bias-corrected estimate SIMEX.
\end{enumerate} 
SIMEX is shown to be asymptotically unbiased and efficient in logistic regression. The statistical model of SIMEX is
\begin{eqnarray}
Z_i=X_i+\sigma\epsilon_i ,i=1,2,\cdots ,N,\nonumber
\end{eqnarray}
where $N$ is the sample size, $\sigma$ is known or can be efficiently estimated and  measurement error $\epsilon_i\sim N(0,1)$. Parameter of interest is estimated by $\hat{\theta}_T$ where
$\hat{\theta}_T=f(\{D_i,W_i,X_i\}_{i=1}^{N})$ is a function of disease status, confounding factors and true exposure.
Naive estimator
$\hat{\theta}_N=f(\{D_i,W_i,Z_i\}_{i=1}^{N})$ is a function of disease status, confounding factors and FFQ measurement observations.
This representation allows us to apply the simulation-based method on any parameter of interest. It is free from assumptions on the joint distribution of $(X,Z)$. Extra measurement error is added through
\begin{eqnarray}
Z_{b,i}(\tau)=Z_i+\tau^{1/2} \sigma \epsilon_{b,i}, b=1,2,\cdots,B;i=1,2,\cdots,N, \nonumber
\end{eqnarray}
where $B$ is the number of times we implement the simulation procedure, $\tau$ is the magnitude of extra measurement error ($\tau>0$) and $\epsilon_{b,i}\sim N(0,1)$ are independent of each other and independent of data. 
An estimate based upon disturbed main data:
$\hat{\theta}_b(\tau)=f(\{D_i,W_i,Z_{b,i}(\tau)\}_{i=1}^{N})$.
For each value of $\tau$, the estimate is $\hat{\theta}(\tau)=\frac{1}{B}\sum_{b=1}^B \hat{\theta}_b(\tau)$.
If we vary the value of $\tau$, we will have a set of estimates and the relationship between $\tau$ and $\hat{\theta}(\tau)$. By extrapolating the relation to $\tau=-1$, the estimate $\hat{\theta}(-1)$ is SIMEX estimator $\hat{\theta}_{SIMEX}$. 

The asymptotic bias of SIMEX estimator is $O(\sigma^4)$ for linear extrapolation. It is better than naive estimator since the asymptotic bias is $O(\sigma^2)$ for naive estimator. Quadratic extrapolation is preferred over linear extrapolation. SIMEX estimator $\hat{\theta}_{SIMEX}$ is asymptotically Gaussian and approximately consistent, which implies that the estimator converges in probability to a value close to true value as sample size grows to infinity \citep{b18}. If the extrapolation function is exact and the variance of measurement error $\sigma^2_{\epsilon}$ is known, SIMEX estimator is exactly consistent \citep{b47}.

\citet{b47} explores the asymptotic properties of SIMEX estimator $\hat{\theta}_{SIMEX}$ by using the association between SIMEX estimator and jackknife estimator with sample size equal to one. The association is explained with jackknife extrapolation. Leave-$(-\infty)$-out extrapolation under jackknife framework is similar to $\tau=-1$ extrapolation in SIMEX procedure. Same as in \citet{b16}, SIMEX assumes that measurement error variance is known or can be efficiently estimated. 
Define
$\Delta(\tau)=\hat{\theta}_b(\tau)-\hat{\theta}(\tau)$,
and it reaches the conclusion that
\begin{eqnarray}
\mbox{Var}(\hat{\theta}_{SIMEX})=-\lim_{\tau\rightarrow -1} \mbox{Var}(\Delta(\tau)).\nonumber
\end{eqnarray}
Jackknife variance estimate of SIMEX estimator includes the following steps. 
\begin{enumerate}
	\item For each value of $\tau$, compute the sample variance $S^2_{\Delta}(\tau)$ of simulated data. $\hat{\theta}_b(\tau),b=1,2,\cdots,B$.
	\item Extrapolate the relation between $S^2_{\Delta}(\tau)$ and $\tau$ to the case where $\tau=-1$. 
	\item Then $S^2_{\Delta}(-1)$ is a variance estimate of SIMEX estimator $\hat{\theta}_{SIMEX}$.
\end{enumerate}

\subsection{Nonparametric Methods}
\label{nonpar-methods}

Nonparametric methods can be applied to approximate the true exposure distribution free from contamination of measurement error.
For additive measurement error, $Z=X+\epsilon$.
Denote the distribution of true exposure $X$ by $g$, distribution of observed exposure $Z$ by $f$ and distribution of measurement error $\epsilon$ by $h$. 
\citet{b45} explains the deconvolution method to derive true exposure distribution $g$ from  $f$ and $h$. 
It is assumed that measurement error distribution $h$ is known or can be efficiently estimated. It integrates nonparametric density estimation into deconvolution of distribution to obtain a density estimate of true exposure distribution $g$.

Nonparametric regression may be applied to logistic regression $D\sim X,W$, linear regression $X\sim Z$ or both \citep{b29}. Nonparametric approximation may be used to estimate likelihood function and score function. An estimator \citep{b14} is solved iteratively from an estimating equation constructed by equaling mixed score function to zero:
\begin{eqnarray}
\sum_{j=1}^{n_1}l(d_j,x_j,\beta)+\sum_{i=n_1+1}^n H_n(d_i,z_i,\beta)=0,\nonumber
\end{eqnarray}
where $n_1$ is the sample size of validation study and $n-n_1$ is the sample size of primary study. The first term is based on the score function of logistic regression in validation study and the second term is from nonparametric approximation of density $p(D|Z)$. 
The advantages of nonparametric methods are as follows \citep{b14}.
\begin{enumerate}
	\item It imposes no parametric assumption on the joint distribution of $(Z,X)$.
	\item It allows nonlinear components to be included in regression.
	\item It considers heteroscedasticity of measurement error. 
\end{enumerate} 
More nonparametric or semi-parametric 
measurement error models are developed in recent years. A nonparametric mixture model is suggested in \citet{b7}. A Bayesian regression of splines is proposed in \citet{b8}.

\subsection{Other Models}
\label{other-models}

\citet{b10} describes another approach for making bias-corrections: hierarchical model and Gibbs sampling. Measurement error model is decomposed into three sub-models: disease model of disease status conditional upon true exposure and confounding factors ($D|X,W,\beta$), measurement model of observed exposure conditional upon true exposure ($Z|X,\lambda$) and exposure model of true exposure conditional upon confounding factors ($X|W,\pi$), where $\beta$, $\lambda$ and $\pi$ are unknown parameters in the sub-models. It assumes that conditional independence assumption holds so that the prior distributions of $(\beta,\lambda,\pi)$ times the joint distribution of $(Z,X,D)|W$ can be written as
\begin{eqnarray}
f(\beta)f(\lambda)f(\pi)\Pi_i(X_i|W_i,\pi)\Pi_i(Z_i|X_i,\lambda)\Pi_i(D_i|X_i,W_i,\beta),\nonumber
\end{eqnarray}
from which we can derive the conditional distributions used in Gibbs sampling. 
Bias-corrected estimator can be derived based upon the posterior distribution.
We require the presence of primary data,  validation data and reliability data to ensure model identifiability. Hierarchical model can be applied to both measurement error model and Berkson error model \citep{b27}. In logistic regression, we impose the following assumptions.
\begin{enumerate}
	\item $(X_i|W_i,\pi)\sim N(x,\sigma^2_x)$, where $N(x,\sigma^2_x)$ is Gaussian distribution with mean $x$ and variance $\sigma^2_x$.
	\item $(Z_i|X_i,\lambda)\sim N(\alpha^\prime +\lambda X_i,\sigma^2_{\epsilon})$.
	\item $(D_i|X_i,W_i,\beta)\sim \mbox{Bernoulli}(\exp(\alpha+W_i\beta_1+X_i\beta_2)/\{1+\exp(\alpha+W_i\beta_1+X_i\beta_2)\})$, where $\mbox{Bernoulli}(p)$ is Bernoulli distribution with success probability $p$.
\end{enumerate}  
Unknown parameters are $(\alpha^\prime,\alpha,\beta_1,\beta_2,\lambda,\sigma^2_x,\sigma^2_\epsilon,x)$. Non-informative priors may be imposed upon these unknown parameters.

Instrumental variable method is applicable in cases where there is another independent measurement $S$ of true exposure $X$ in the study \citep{b48}. 
Denote $Z$ as the observed predictor of $X$ in the study and $D$ is the disease status.
Instrumental variable satisfies the following conditions.
\begin{enumerate}
	\item $\mbox{Cor}(S,X)\neq 0$, $\mbox{Cor}(S,Z-X)=0$.
	\item $\mbox{Cor}(S,D-\mbox{E}(D|X))=0$, where $\mbox{Cor}(a,b)$ is the correlation between $a$ and $b$.
	\item In addition, \citet{b48} assumes that $\mbox{E}(S|X)=X$ and $\mbox{Var}(S|X)=\mbox{Var}(Z|X)$.
\end{enumerate} 
For linear regression, bias-corrected estimator based on data $(D,Z,S)$ is represented in a nice closed form. For generalized linear regression, bias-corrected estimator is derived through linear approximations of the regression model \citep{b48}. Bias-corrected estimator is approximately consistent.

\section{Simulation}
\label{simulation}

In nutritional epidemiology, researchers are interested in the relation between disease status and dietary habits of people. Dietary intake measured with food frequency questionnaire (FFQ) is subject to substantial random and systematic measurement error. Measurement error introduces bias into the estimated effect of dietary intake on disease status, leads to narrow confidence interval on parameter of interest and power loss in hypothesis test. This is why we need bias-correction methods to develop better estimates in the presence of measurement error. The simulation study is conducted for the comparison of various bias-correction methods applicable in nutritional epidemiology studies.

Data in nutritional epidemiology contains FFQ, reference measurement data including 24-hour recall (24HR), 7-day dairy (7DD) and food record (FR), and biomarker measurement data \citep[chap.~4]{b41}. Denote measurements in FFQ by $Z_{ij}$, reference measurements by $F_{ij}$ and biomarker measurements by $M_{ij}$, $i=1,2,\cdots,N$ and $j=1,2,\cdots,J$, where $N$ is the number of subjects enrolled and $J$ is the number of timestamps in the validation study. We also denote $X_i$ to be the true exposure of the $i$th individual. 
Logistic regression \citep{b1,b3} is applied to model the relation between disease status $D$ and true exposure $X$:
\begin{eqnarray}
P(D=1|X)=\exp(\alpha+\beta X)/\{1+\exp(\alpha+\beta X)\},\nonumber
\end{eqnarray}
where $D$ is binary and takes values from $\{0,1\}$, $\alpha$ is the unknown intercept parameter and $\beta$ is the unknown slope parameter of interest in nutritional epidemiology study.
A typical measurement error model in  nutritional epidemiology \citep{b3,b51} is posited to be the following form:
\begin{eqnarray}
Z_{ij}&=&\mu_{Zj}+\lambda_{Z0}+\lambda_{Z1} X_i+r_i+\varepsilon_{Zij},\nonumber\\
F_{ij}&=&\mu_{Fj}+\lambda_{F0}+\lambda_{F1} X_i+s_i+\varepsilon_{Fij},\nonumber\\
M_{ij}&=&\mu_{Mj}+X_i+\varepsilon_{Mij},\nonumber
\end{eqnarray}
where $\mu_{Zj}$ is the time-specific bias of $Z_{ij}$ and $\sum_{j=1}^J \mu_{Zj}=0$, $\mu_{Fj}$ is the time-specific bias of $F_{ij}$ and $\sum_{j=1}^J \mu_{Fj}=0$, $\mu_{Mj}$ is the time-specific bias of $M_{ij}$ and $\sum_{j=1}^J \mu_{Mj}=0$, $(\lambda_{Z0},\lambda_{Z1},\lambda_{F0},\lambda_{F1})$ are the unknown parameters, $r_i\sim (r,\sigma^2_r)$ is the systematic within-subject measurement error in $Z_{ij}$, $\varepsilon_{Zij}\sim (0,\sigma^2_{Zi})$ is the random within-subject measurement error in $Z_{ij}$, $\sigma^2_{Zi}$ is simulated from uniform distribution centered at $\sigma^2_Z$, $s_i\sim (s,\sigma^2_s)$ is the systematic within-subject measurement error in $F_{ij}$, $\varepsilon_{Fij}\sim (0,\sigma^2_{Fi})$ is the random within-subject measurement error in $F_{ij}$, $\varepsilon_{Mij}\sim (0,\sigma^2_M)$ is the random within-subject measurement error in $M_{ij}$. True exposure $X_i$ is simulated from Gaussian distribution $N(x,\sigma^2_x)$.  
\citet{b3} simplifies the model to be
\begin{eqnarray}
Z_{ij}&=&\mu_{Zj}+\lambda_{Z0}+\lambda_{Z1} X_i+r_i+\varepsilon_{Zij},\nonumber\\
M_{ij}&=&\mu_{Mj}+X_i+\varepsilon_{Mij},\nonumber
\end{eqnarray}
where the redundant measurement is removed. Measurement errors in the biomarker measurements are independent of those in FFQ. Performances of naive estimator (Naive), regression calibration (RC) estimator and  likelihood approximation estimator (Lik) are studied in the simulation with this model.

\subsection{Factors}
\label{factors-importance}

Denote the sample size of primary study i.e. FFQ to be $n_2$ and the sample size of validation study to be $n_1$. In nutritional epidemiology, $n_2\gg n_1$ and we set $n_2$ to be $100,000$ and $n_1$ to be $1,000$ in our simulation. Intuitively, as ratio $n_1/n_2$ increases, these bias-correction methods should all perform better. Two levels of slope parameter $\beta$ (related to the disease odds ratio) are considered in the simulation. Measurement error distribution is examined since estimators are derive under the normality assumption, whereas in nutritional epidemiology measurement error distribution is usually skewed. Whether measurement error variance $\sigma^2_{Zi}$ is constant across $i$ or not is pertinent because homoscedasticity of measurement error is assumed in most measurement error models. 

\subsection{Data Generation}
\label{data-generation}

There are two choices of parameter: (i) $\alpha=-1$, $\beta=0.1$ and (ii) $\alpha=-1$, $\beta=1$, and two different distributions of random measurement error: (i) $N(0,\sigma^2_\varepsilon)$ and (ii) $\{\mbox{lognormal}(0,1)-\sqrt{e}\}\sigma_\varepsilon/\sqrt{(e-1)e}$.  The second distribution is transformed from lognormal distribution. After transformation, it is a skewed distribution with mean $0$ and variance $\sigma^2_\varepsilon$. Taking all combinations of these two factors, we form four sets of conditions for the simulation, which are (i) $\alpha=-1$, $\beta=1$, $N(0,\sigma^2_\varepsilon)$; (ii) $\alpha=-1$, $\beta=1$, $\{\mbox{lognormal}(0,1)-\sqrt{e}\}\sigma_\varepsilon/\sqrt{(e-1)e}$; (iii) $\alpha=-1$, $\beta=0.1$, $N(0,\sigma^2_\varepsilon)$ and (iv) $\alpha=-1$, $\beta=0.1$, $\{\mbox{lognormal}(0,1)-\sqrt{e}\}\sigma_\varepsilon/\sqrt{(e-1)e}$. 

Other relevant parameters are set as $\sigma^2_{Z}/\sigma^2_x=0.3$, $\sigma^2_Z=0.3$, $\sigma^2_x=1$, $n_1=1,000$, $n_2=100,000$, $\lambda_{Z0}=0.1$, $\lambda_{Z1}=0.5$, $r=0.05$, $\sigma^2_r=0.05$, $x=0$ for cases (i)(iii)(iv) and $x=1$ for case (ii), $\sigma^2_{M}=0.0001$ and $J=2$. Simulation size is $M=1000$.
True exposure $X_i$ is generated as a random sample of size $N=n_1+n_2$ from Gaussian distribution $N(0,1)$. 

For the main dataset of size $n_2$ with  disease status and FFQ nutrient intake measurements, we use the first $n_2$ simulated true exposure. Generate a scalar from uniform distribution $\mbox{Uniform}(0,1)$. For scalar less than $\exp(\alpha+\beta X_i)/\{1+\exp(\alpha+\beta X_i)\}$, we set disease status to be $1$. Otherwise we set disease status to be $0$. 

For FFQ nutrient intake in the main data, first we generate individual measurement error variances $\sigma^2_{Zi}$ from uniform distribution $\mbox{Uniform}(\sigma^2_Z-0.05,\sigma^2_Z+0.05)$. Random measurement error in $Z_i$ is simulated from transformed lognormal distribution or normal distribution with mean $0$ and variance $\sigma^2_{Zi}$. Systematic measurement error in $Z_i$ is generated from $N(r,\sigma^2_r)$. Then we add up $\lambda_{Z0}+\lambda_{Z1}X_i$, systematic measurement error and random measurement error and obtain simulated observed exposure in FFQ. 

As for the validation dataset of size $n_1$ with nutrient intake from FFQ and biomarker measurements repeated at $J$ timestamps,  individual measurement error variance $\sigma^2_{Zi}$ is generated from uniform distribution $\mbox{Uniform}(\sigma^2_Z-0.05,\sigma^2_Z+0.05)$. Random measurement error $\varepsilon_{Zij}$ is generated from Gaussian distribution $N(0,\sigma^2_{Zi})$ or transformed lognormal distribution with mean $0$ and variance $\sigma^2_{Zi}$. Systematic measurement error $r_i$ is generated from Gaussian distribution $N(r,\sigma^2_r)$. Time-specific bias $\mu_{Zj},j=1,2,\cdots,J-1$ are independently generated from uniform distribution $\mbox{Uniform}(-0.01,0.01)$ and $\mu_{ZJ}=-\sum_{j=1}^{J-1} \mu_{Zj}$. Then we add up $\lambda_{Z0}+\lambda_{Z1}X_i$, systematic measurement error, random measurement error and time-specific bias and obtain simulated observed exposure in the validation study. 

As for biomarker measurements in the validation data, individual random measurement error variance $\sigma^2_{Mi}$ is generated from uniform distribution $\mbox{Uniform}(0,2\sigma^2_M)$. Random measurement error $\varepsilon_{Mij}$ is generated from Gaussian distribution $N(0,\sigma^2_{Mi})$ or transformed lognormal distribution with mean $0$ and variance $\sigma^2_{Mi}$. Time-specific bias $\mu_{Mj},j=1,2,\cdots,J-1$ are independently generated from uniform distribution $\mbox{Uniform}$ $(-0.001,0.001)$ and $\mu_{MJ}=-\sum_{j=1}^{J-1} \mu_{Mj}$. Biomarker measurement is the sum of true exposure $X_i$, random measurement error and time-specific bias.

\subsection{Estimators}
\label{estimators-variances}

Naive estimator of $\beta$ is constructed by fitting logistic regression of $D_i$ on $Z_i$ in the primary data. Slope estimate $\hat{\beta}_N$ is naive estimator. Variance of naive estimator is reported in the regression output.
Fit a linear regression of $M_{ij}$ on $Z_{ij}$ and the estimated slope is $\hat{\lambda}$. RC estimator of $\beta$ is $\hat{\beta}_R=\hat{\beta}_N/\hat{\lambda}$. Variance of RC estimator is 
$Var(\hat{\beta}_R)=(1/\hat{\lambda}^2)Var(\hat{\beta}_N)+(\hat{\beta}_N^2/\hat{\lambda}^4)Var(\hat{\lambda})$ \citep{b1}. Likelihood-based estimator proposed in \citet{b1} is also studied in the simulation.

\subsection{Results and Interpretation}
\label{results-interpretation}

Performance indicators in the tables of simulation results are as follows.
\begin{enumerate}
	\item "Avg est" is the average of all estimates in the simulation.
	\item "Avg se" is the average of all standard error estimates in the simulation.
	\item "MC sd" is the standard deviation of all estimates in the simulation.
	\item "Coverage" is the proportion of $95\%$ confidence intervals constructed with Gaussian approximation that cover true parameter. It indicates whether standard error estimate is reasonable.
\end{enumerate}

\subsubsection{$\beta=1$ and Gaussian Measurement Error}

Simulation results for the case where $\beta=1$ and measurement error distribution is Gaussian are displayed in table \ref{t1}.
\begin{table}[htpb]\centering
		\caption{Performance of bias-correction methods under $\beta=1$ and Gaussian measurement error.}
	\label{t1}
		\begin{tabular}{@{}cccccc@{}}
			\toprule
			method & Avg est & Avg se & MC sd & ratio of Avg se to MCsd & coverage \\ \midrule
			Naive & 7.49E-01 & 9.80E-03 & 9.78E-03 & 1.00E+00 & 0.00 \\
			&(3.09E-04)&(1.02E-06)&(2.15E-04)&(2.21E-02)&(0.00)\\
			RC & 9.00E-01 & 2.66E-02 & 3.24E-02 & 8.21E-01 & 0.09 \\
			&(1.03E-03)&(5.37E-05)&(8.07E-04)&(2.03E-02)&(9.05E-03)\\
			Lik & 9.67E-01 & 3.28E-02 & 3.95E-02 & 8.32E-01 & 0.76 \\ 
			&(1.25E-03)&(8.14E-05)&(9.83E-04)&(2.05E-02)&(1.35E-02)\\\bottomrule
		\end{tabular}
\end{table}

Likelihood-based estimator contains the least bias and is closest to the true value of $\beta$. The $95\%$ confidence intervals formed by est.$\pm$ $1.96$se. have coverage proportions of only $0$ and $0.09$ for naive and RC estimators. The standard error of likelihood-based estimator is closest to the reasonable standard error. The standard errors of naive estimator and RC estimator are severely underestimated. Ratios of the average standard error to Monte Carlo standard deviation are all close to $1$. 

When true odds ratio ($\exp(\beta)$) is large, none of these methods shows good coverage close to $95\%$. Point estimates from RC and  likelihood-based estimation are acceptably close to the true value. Naive estimate is obviously attenuated towards $0$ and is less than the average estimate of RC method and likelihood-based method. 

\subsubsection{$\beta=1$ and Skewed Measurement Error}

Simulation results of the case where $\beta=1$ and measurement error distribution is skewed are presented in table \ref{t2}.
\begin{table}[htpb]\centering
		\caption{Performance of bias-correction methods under $\beta=1$ and skewed measurement error.}
	\label{t2}
		\begin{tabular}{@{}cccccc@{}}
			\toprule
			method & Avg est & Avg se & MC sd & ratio of Avg se to MC sd & coverage \\ \midrule
			Naive & 7.61E-01 & 1.03E-02 & 1.29E-02 & 7.96E-01 & 0.00 \\
			&(4.08E-04)&(1.77E-06)&(2.76E-04)&(1.71E-02)&(0.00)\\
			RC & 9.11E-01 & 2.71E-02 & 1.05E-01 & 2.58E-01 & 0.23 \\
			&(3.33E-03)&(1.82E-04)&(1.01E-02)&(2.57E-02)&(1.33E-02)\\
			Lik & 1.00E+00 & 6.31E-02 & 6.53E-01 & 9.67E-02 & 0.42 \\ 
			&(2.06E-02)&(2.88E-02)&(5.16E-01)&(1.55E-01)&(1.56E-02)\\\bottomrule
		\end{tabular}
\end{table}

The average estimate bias is comparable to table \ref{t1} for each estimator. The average standard errors and Monte Carlo standard deviations are greater than those in table \ref{t1}. More outliners are present in generated exposure factors when we use shifted and scaled lognormal distribution instead of Gaussian distribution for measurement error. Ratios of the average standard error to Monte Carlo standard deviation are much less than $1$ and much lower than those in table \ref{t1}. This phenomenon indicates that the standard error formulas underestimate variability in this case and Monte Carlo standard deviation is a more reliable estimation of variability. 

\subsubsection{$\beta=0.1$ and Gaussian Measurement Error}

Simulation results when $\beta=0.1$ and when measurement error distribution is skewed are given in table \ref{t3}.
\begin{table}[htpb]\centering
		\caption{Performance of bias-correction methods under $\beta=0.1$ and Gaussian measurement error.}
	\label{t3}
		\begin{tabular}{@{}cccccc@{}}
			\toprule
			method & Avg est & Avg se & MC sd & ratio of Avg se to MC sd & coverage \\ \midrule
			Naive & 8.32E-02 & 9.21E-03 & 9.46E-03 & 9.74E-01 & 0.56 \\
			&(2.99E-04)&(7.91E-07)&(2.11E-04)&(2.18E-02)&(1.57E-02)\\
			RC & 9.98E-02 & 1.14E-02 & 1.19E-02 & 9.58E-01 & 0.93 \\
			&(3.75E-04)&(1.27E-05)&(2.63E-04)&(2.12E-02)&(7.96E-03)\\
			Lik & 9.99E-02 & 1.14E-02 & 1.19E-02 & 9.58E-01 & 0.93 \\ 
			&(3.76E-04)&(1.28E-05)&(2.64E-04)&(2.12E-02)&(8.02E-03)\\\bottomrule
		\end{tabular}
\end{table}

In this case, true odds ratio of infection $D=1$ versus no infection $D=0$ is $\exp(0.1)=1.105$, less than first two cases where true odds ratio is $\exp(1)=2.718$. RC estimator and likelihood-based estimator are similar. Both show little bias and standard errors are reasonable with coverage proportion close to $95\%$. RC estimator and likelihood-based estimator both perform well for rare disease and Gaussian measurement error. 

Ratios of the average standard error to Monte Carlo standard deviation are all close to $1$. Coverage of naive estimate is $0.56$ much smaller compared to coverage of RC estimate and likelihood-based estimate. Attenuation effect is obvious here since the average of naive estimates is $0.83$, farther away from true value $1$ than RC estimate and likelihood-based estimate. The standard error of naive estimate is still underestimated in this case, but much better than in the former two cases, where coverage is $0$. 

\subsubsection{$\beta=0.1$ and Skewed Measurement Error}

Simulation results where $\beta=0.1$ and measurement error distribution is skewed are summarized in table \ref{t4}.
\begin{table}[htpb]\centering
		\caption{Performance of bias-correction methods under $\beta=0.1$ and skewed measurement error.}
	\label{t4}
		\begin{tabular}{@{}cccccc@{}}
			\toprule
			method & Avg est & Avg se & MC sd & ratio of Avg se to MC sd & coverage \\ \midrule
			Naive & 8.13E-02 & 9.02E-03 & 8.73E-03 & 1.03E+00 & 0.46 \\
			&(2.76E-04)&(2.51E-06)&(1.97E-04)&(2.33E-02)&(1.58E-02)\\
			RC & 9.78E-02 & 1.12E-02 & 1.59E-02 & 7.02E-01 & 0.86 \\
			&(5.03E-04)&(4.57E-05)&(8.07E-04)&(3.47E-02)&(1.11E-02)\\
			Lik & 9.78E-02 & 1.12E-02 & 1.60E-02 & 7.01E-01 & 0.86 \\ 
			&(5.05E-04)&(4.63E-05)&(8.17E-04)&(3.50E-02)&(1.11E-02)\\\bottomrule
		\end{tabular}
\end{table}

Performances of RC estimator and likelihood-based estimator are similar when true odds ratio is close to $1$ (rare disease). For skewed measurement error distribution, more outliners are present in the simulation results and standard errors are inflated. Accordingly ratio of the average standard error to Monte Carlo standard deviation is much less than $1$ both for RC estimator and for likelihood-based estimator. 

Coverage proportions of RC and likelihood-based estimators are not as good as those in case $3$ where measurement error distribution is Gaussian. The standard error is underestimated using the formula derived for skewed measurement error distribution. Naturally due to attenuation effect, the average naive estimate is less than RC estimate. The standard error of naive estimate is more severely underestimated than that of RC and likelihood-based estimator.

\subsubsection{Conclusions}

Naive estimator is biased towards null value. The standard error of naive estimator is subject to underestimation. Regression calibration (RC) method and likelihood-based model are applicable when the true disease odds ratio is small and when random measurement error is Gaussian. RC estimate and likelihood-based estimate show similar performance when true odds ratio is small. 

For cases with large true disease odds ratio, we reach the following conclusions.
\begin{enumerate}
	\item RC estimate is worse than likelihood-based method and better than naive estimate.
	\item Estimation from likelihood-based method is close to the true value but standard error is underestimated.
	\item Naive estimator is worse in terms of both bias and standard error of estimator.
	\item Bias of RC estimate is slightly greater than likelihood-based estimate.
	\item Coverage proportion of RC estimate is much worse than likelihood-based estimate. 
\end{enumerate}

For skewed measurement error distribution, we reach the following conclusions.
\begin{enumerate}
	\item Likelihood-based bias-corrected estimator is better than RC estimator, which is better than naive estimator.
	\item Coverage proportion of these methods are worse, which implies that standard errors face more underestimation than Gaussian cases.
	\item Average estimate for these three methods are biased more towards null value than Gaussian cases.
	\item Monte Carlo standard errors of RC estimate and likelihood-based estimate are greater.
	\item Ratio of average standard error to Monte Carlo standard deviation is much less than $1$ and close to $0$.
	\item Monte Carlo standard deviation of the naive estimate is still close to the average standard error and is slightly less than that of the Gaussian case.
\end{enumerate}  
This is different from RC estimate and likelihood-based estimate since the standard error formulas of RC estimate and likelihood-based estimate are derived with Delta theorem but the standard error formula of naive estimate is derived from linear model.

\section{Discussion}
\label{discussion}

Regression calibration method is widely used for correcting bias induced by measurement error in nutritional epidemiology. Other methods are not as frequently used. Methods such as SIMEX, EM and MIME require that measurement error variance should be known or can be efficiently estimated. SIMEX and MIME measurement error models are only applicable to random measurement error and do not consider systematic measurement error. Performances of SIMEX and MIME depend upon the ratio of validation sample size to primary study sample size. As the proportion of validated individuals increases, performance of bias-correction is better.

Proper choice of measurement error bias-correction models depends upon data availability in nutritional epidemiology.
Methods such as regression calibration,  hierarchical models and likelihood-based methods integrate large primary study and small-scale precise validation study to conduct analysis. Bayesian methods require additionally the reliability study to ensure model identifiability. Instrumental variable method considers the situation where a second independent measurement of the same true exposure is available. This method is only applicable to this specific type of data in nutritional epidemiology.
In addition, computational feasibility of bias-correction method is also of concern. 

More complex correlation structure may be incorporated into measurement error models. Differential measurement error may be considered. Measurement errors in different exposure factors may be correlated. Measurement error may be correlated with disease status. Heteroscedasticity in measurement error is of interest as well.
For application in nutritional epidemiology, 
researchers may add interaction terms between individual covariates and personal nutritional intake to linear multiple regression or logistic regression. Through bias-correction, unbiased hypothesis testing results can be applied to select a proper association model between nutritional intake exposures and disease status.

\bibliographystyle{unsrt}


\begin{thebibliography}{38}
	\providecommand{\natexlab}[1]{#1}
	\providecommand{\url}[1]{\texttt{#1}}
	\expandafter\ifx\csname urlstyle\endcsname\relax
	\providecommand{\doi}[1]{doi: #1}\else
	\providecommand{\doi}{doi: \begingroup \urlstyle{rm}\Url}\fi
	
	\bibitem[Armstrong et~al.(1989)Armstrong, Whittemore, and Howe]{b13}
	BG~Armstrong, AS~Whittemore, and GR~Howe.
	\newblock Analysis of case-control data with covariate measurement error -
	application to diet and colon cancer.
	\newblock \emph{Statistics in Medicine}, 8\penalty0 (9):\penalty0 1151--1163,
	1989.
	
	\bibitem[Berry et~al.(2002)Berry, Carroll, and Ruppert]{b8}
	SM~Berry, RJ~Carroll, and D~Ruppert.
	\newblock Bayesian smoothing and regression splines for measurement error
	problems.
	\newblock \emph{Journal of the American Statistical Association}, 97\penalty0
	(457):\penalty0 160--169, 2002.
	
	\bibitem[Carroll(1989)]{b34}
	RJ~Carroll.
	\newblock Covariance analysis in generalized linear measurement error models.
	\newblock \emph{Statistics in Medicine}, 8\penalty0 (9):\penalty0 1075--1093,
	1989.
	
	\bibitem[Carroll and Stefanski(1990)]{b33}
	RJ~Carroll and LA~Stefanski.
	\newblock Approximate quasi-likelihood estimation in models with surrogate
	predictors.
	\newblock \emph{Journal of the American Statistical Association}, 85\penalty0
	(411):\penalty0 652--663, 1990.
	
	\bibitem[Carroll and Wand(1991)]{b14}
	RJ~Carroll and MP~Wand.
	\newblock Semiparametric estimation in logistic measurement error models.
	\newblock \emph{Journal of the Royal Statistical Society Series
		B-Methodological}, 53\penalty0 (3):\penalty0 573--585, 1991.
	
	\bibitem[Carroll et~al.(1996)Carroll, Kuchenhoff, Lombard, and Stefanski]{b18}
	RJ~Carroll, H~Kuchenhoff, F~Lombard, and LA~Stefanski.
	\newblock Asymptotic for the simex estimator in nonlinear measurement error
	models.
	\newblock \emph{Journal of the American Statistical Association}, 91\penalty0
	(433):\penalty0 242--250, 1996.
	
	\bibitem[Cole et~al.(2006)Cole, Chu, and Greenland]{b39}
	S~Cole, H~Chu, and S~Greenland.
	\newblock Multiple-imputation for measurement-error correction.
	\newblock \emph{international Journal of Epidemiology}, 35\penalty0
	(4):\penalty0 1074--1081, 2006.
	
	\bibitem[Cook and Stefanski(1994)]{b16}
	JR~Cook and LA~Stefanski.
	\newblock Simulation-extrapolation estimation in parametric measurement error
	models.
	\newblock \emph{Journal of the American Statistical Association}, 89\penalty0
	(428):\penalty0 1314--1328, 1994.
	
	\bibitem[Delaigle et~al.(2006)Delaigle, Hall, and Qiu]{b37}
	A~Delaigle, P~Hall, and P~Qiu.
	\newblock Nonparametric methods for solving the berkson errors-in-variables
	problem.
	\newblock \emph{Journal of the Royal Statistical Society: Series B},
	68\penalty0 (2):\penalty0 201--220, 2006.
	
	\bibitem[Greenwood(2012)]{b40}
	DarrenC. Greenwood.
	\newblock Measurement errors in epidemiology.
	\newblock In Yu-Kang Tu and Darren~C. Greenwood, editors, \emph{Modern Methods
		for Epidemiology}, chapter~3, pages 33--55. Springer Netherlands, 2012.
	
	\bibitem[Haukka(1995)]{b32}
	JK~Haukka.
	\newblock Correction for covariate measurement error in generalized
	linear-models - a bootstrap approach.
	\newblock \emph{Biometrics}, 51\penalty0 (3):\penalty0 1127--1132, 1995.
	
	\bibitem[Jenab et~al.(2009)Jenab, Slimani, Bictash, and al]{b25}
	M~Jenab, N~Slimani, M~Bictash, and et~al.
	\newblock Biomarkers in nutritional epidemiology: Applications, needs and new
	horizons.
	\newblock \emph{Human Genetics}, 125\penalty0 (5-6):\penalty0 507--525, 2009.
	
	\bibitem[Kipnis et~al.(1997)Kipnis, Freedman, Brown, and al]{b6}
	V~Kipnis, LS~Freedman, CC~Brown, and et~al.
	\newblock Effect of measurement error on energy-adjustment models in
	nutritional epidemiology.
	\newblock \emph{American Journal of Epidemiology}, 146\penalty0 (10):\penalty0
	842--855, 1997.
	
	\bibitem[Kipnis et~al.(2001)Kipnis, Midthune, Freedman, and al]{b22}
	V~Kipnis, D~Midthune, LS~Freedman, and et~al.
	\newblock Empirical evidence of correlated biases in dietary assessment
	instruments and its implications.
	\newblock \emph{American Journal of Epidemiology}, 153\penalty0 (4):\penalty0
	394--403, 2001.
	
	\bibitem[Kipnis et~al.(2002)Kipnis, Midthune, Freedman, and al]{b23}
	V~Kipnis, D~Midthune, L~Freedman, and et~al.
	\newblock Bias in dietary-report instruments and its implications for
	nutritional epidemiology.
	\newblock \emph{Public Health Nutrition}, 5\penalty0 (6A):\penalty0 915--923,
	2002.
	
	\bibitem[Kipnis et~al.(2003)Kipnis, Subar, Midthune, and al]{b3}
	V~Kipnis, AF~Subar, D~Midthune, and et~al.
	\newblock Structure of dietary measurement error: Results of the open biomarker
	study.
	\newblock \emph{American Journal of Epidemiology}, 158\penalty0 (1):\penalty0
	14--21, JUL 2003.
	
	\bibitem[Madansky(1959)]{b4}
	A~Madansky.
	\newblock The fitting of straight-lines when both variables are subject to
	error.
	\newblock \emph{Journal of the American Statistical Association}, 54\penalty0
	(285):\penalty0 173--205, 1959.
	
	\bibitem[Michels et~al.(2004)Michels, Bingham, Luben, and al]{b20}
	KB~Michels, SA~Bingham, R~Luben, and et~al.
	\newblock The effect of correlated measurement error in multivariate models of
	diet.
	\newblock \emph{American Journal of Epidemiology}, 160\penalty0 (1):\penalty0
	59--67, 2004.
	
	\bibitem[Richardson and Gilks(1993)]{b10}
	S~Richardson and WR~Gilks.
	\newblock Conditional-independence models for epidemiologic studies with
	covariate measurement error.
	\newblock \emph{Statistics in Medicine}, 12\penalty0 (18):\penalty0 1703--1722,
	1993.
	
	\bibitem[Roeder et~al.(1996)Roeder, Carroll, and Lindsay]{b7}
	K~Roeder, RJ~Carroll, and BG~Lindsay.
	\newblock A semiparametric mixture approach to case-control studies with errors
	in covariables.
	\newblock \emph{Journal of the American Statistical Association}, 91\penalty0
	(434):\penalty0 722--732, 1996.
	
	\bibitem[Rosner and Gore(2001)]{b38}
	B~Rosner and R~Gore.
	\newblock Measurement error correction in nutritional epidemiology based on
	individual foods, with application to the relation of diet to breast cancer.
	\newblock \emph{American Journal of Epidemiology}, 154\penalty0 (9):\penalty0
	827--835, 2001.
	
	\bibitem[Rosner et~al.(1989)Rosner, Willett, and Spiegelman]{b1}
	B~Rosner, WC~Willett, and D~Spiegelman.
	\newblock Correction of logistic-regression relative risk estimates and
	confidence-intervals for systematic within-person measurement error.
	\newblock \emph{Statistics in Medicine}, 8\penalty0 (9):\penalty0 1051--1069,
	Sep 1989.
	
	\bibitem[Rosner et~al.(2008)Rosner, Michels, Chen, and Day]{b51}
	B~Rosner, KB~Michels, YH~Chen, and NE~Day.
	\newblock Measurement error correction for nutritional exposures with
	correlated measurement error: use of the method of triads in a longitudinal
	setting.
	\newblock \emph{Statist. Med}, 27\penalty0 (18):\penalty0 3466--89, 2008.
	
	\bibitem[Schafer(1987)]{b9}
	DW~Schafer.
	\newblock Covariate measurement error in generalized linear-models.
	\newblock \emph{Biometrika}, 74\penalty0 (2):\penalty0 385--391, 1987.
	
	\bibitem[Sinha et~al.(2010)Sinha, Mallick, Kipnis, and Carroll]{b29}
	S~Sinha, BK~Mallick, V~Kipnis, and RJ~Carroll.
	\newblock Semiparametric bayesian analysis of nutritional epidemiology data in
	the presence of measurement error.
	\newblock \emph{Biometrics}, 66\penalty0 (2):\penalty0 444--454, 2010.
	
	\bibitem[Song et~al.(2010)Song, Lawson, and Nitcheva]{b27}
	HR~Song, AB~Lawson, and D~Nitcheva.
	\newblock Bayesian hierarchical models for food frequency assessment.
	\newblock \emph{Canadian Journal of Statistics-Revue Canadienne De
		Statistique}, 38\penalty0 (3):\penalty0 506--516, 2010.
	
	\bibitem[Spiegelman(1994)]{b21}
	D~Spiegelman.
	\newblock Cost-efficient study designs for relative risk modeling with
	covariate measurement error.
	\newblock \emph{Journal of Statistical Planning And inference}, 42\penalty0
	(1-2):\penalty0 187--208, 1994.
	
	\bibitem[Spiegelman et~al.(1997)Spiegelman, McDermott, and Rosner]{b11}
	D~Spiegelman, A~McDermott, and B~Rosner.
	\newblock Regression calibration method for correcting measurement-error bias
	in nutritional epidemiology.
	\newblock \emph{American Journal of Clinical Nutrition}, 65\penalty0
	(4):\penalty0 S1179--S1186, 1997.
	
	\bibitem[Stefanski and Buzas(1995)]{b48}
	LA~Stefanski and JS~Buzas.
	\newblock Instrumental variable estimation in binary regression measurement
	error models.
	\newblock \emph{Journal of the American Statistical Association}, 90\penalty0
	(430):\penalty0 541--550, 1995.
	
	\bibitem[Stefanski and Carroll(1985)]{b15}
	LA~Stefanski and RJ~Carroll.
	\newblock Covariate measurement error in logistic-regression.
	\newblock \emph{Annals of Statistics}, 13\penalty0 (4):\penalty0 1335--1351,
	1985.
	
	\bibitem[Stefanski and Carroll(1987)]{b46}
	LA~Stefanski and RJ~Carroll.
	\newblock Conditional scores and optimal scores for generalized linear
	measurement- error models.
	\newblock \emph{Biometrika}, 74\penalty0 (4):\penalty0 703--716, 1987.
	
	\bibitem[Stefanski and Carroll(1990)]{b45}
	LA~Stefanski and RJ~Carroll.
	\newblock Deconvolving kernel density estimators.
	\newblock \emph{Journal of theoretical and Applied Statistics}, 21\penalty0
	(2):\penalty0 169--184, 1990.
	
	\bibitem[Stefanski and Cook(1995)]{b47}
	LA~Stefanski and JR~Cook.
	\newblock Simulation-extrapolation: the measurement error jackknife.
	\newblock \emph{Journal of the American Statistical Association}, 90\penalty0
	(432):\penalty0 1247--1256, 1995.
	
	\bibitem[Thomas et~al.(1993)Thomas, Stram, and Dwyer]{b17}
	D~Thomas, D~Stram, and J~Dwyer.
	\newblock Exposure measurement error: influence on exposure-disease
	relationships and methods of correction.
	\newblock \emph{Annual Review of Public Health}, 14:\penalty0 69--93, 1993.
	
	\bibitem[Tosteson et~al.(2003)Tosteson, Buzas, Demidenko, and Karagas]{b30}
	TD~Tosteson, JS~Buzas, E~Demidenko, and M~Karagas.
	\newblock Power and sample size calculations for generalized regression models
	with covariate measurement error.
	\newblock \emph{Statistics in Medicine}, 22\penalty0 (7):\penalty0 1069--1082,
	2003.
	
	\bibitem[Whittemore and Keller(1988)]{b35}
	AS~Whittemore and JB~Keller.
	\newblock Approximations for regression with covariate measurement error.
	\newblock \emph{Journal of the American Statistical Association}, 83\penalty0
	(404):\penalty0 1057--1066, 1988.
	
	\bibitem[Willett(1989)]{b2}
	W~Willett.
	\newblock An overview of issues related to the correction of non-differential
	exposure measurement error in epidemiologic studies.
	\newblock \emph{Statistics in Medicine}, 8\penalty0 (9):\penalty0 1031--1040,
	Sep 1989.
	
	\bibitem[Willett(2013)]{b41}
	W.~Willett.
	\newblock \emph{Nutritional Epidemiology}.
	\newblock Monographs in Epidemiology and Biostatistics. OUP USA, 2013.
	
\end{thebibliography}

\end{document}